\documentclass[prb,aps,amsmath,preprint]{revtex4}

\usepackage{graphicx}

\usepackage{dcolumn}

\usepackage{bm}

\begin{document}

\title{Density functional study of elastic and vibrational properties of the Heusler-type alloys Fe$_2$VAl and Fe$_2$VGa} 
 
\author{V. Kanchana$^{1*}$, G. Vaitheeswaran$^{1,2}$, 
Yanming Ma$^{3}$, Yu Xie$^{3}$, A. Svane$^{4}$ and O. Eriksson$^{5}$} 
\affiliation{$^{1}$Applied Materials Physics, Department of Materials Science and Engineering, 
Royal Institute of Technology, Brinellv\"agen 23, 100 44 Stockholm, Sweden \\
$^{2}$Advanced Centre of Research in High Energy Materials (ACRHEM), University of Hyderabad, Gachibowli, Hyderabad 500 046, Andhra Pradesh, India\\
$^{3}$National Lab of Superhard Materials, Jilin University, 
Changchun 130012, P. R. China \\
$^{4}$Department of Physics and Astronomy, University of 
Aarhus, DK-8000 Aarhus C, Denmark\\
$^{5}$Department of Physics and Materials Science, Uppsala University, Box 590, 
SE-751 21, Uppsala, Sweden\\}

\date{\today}

\begin{abstract}
The structural and elastic properties as well as phonon-dispersion relations of the Heusler-type alloys Fe$_2$VAl and Fe$_2$VGa   
are computed using density-functional and density-functional perturbation
theory within the generalized-gradient approximation.
The calculated equilibrium lattice constants agree well with the
 experimental values. The elastic constants of Fe$_2$VAl and Fe$_2$VGa are predicted for 
the first time. From the elastic constants the shear modulus, Young's modulus, Poisson's ratio, 
sound velocities and Debye temperatures are obtained. By 
analyzing the ratio between the bulk and shear modulii, we conclude that both  Fe$_2$VAl and Fe$_2$VGa are brittle in nature.
The computed phonon-dispersion relation shows that both compounds are
dynamically stable in the L1$_2$ structure without any imaginary phonon frequencies.
The isomer shifts of Fe in the two compounds are discussed in terms of the Fe s partial density of states, which reveal
larger ionicity/less hybridization in Fe$_2$VGa than in Fe$_2$VAl. For the
 same reason the Cauchy pressure is negative in
Fe$_2$VAl but positive in  Fe$_2$VGa.
\end{abstract}

\maketitle

\section {Introduction} 
The Heusler-type intermetallic compounds Fe$_2$VAl and Fe$_2$VGa
have recently attracted great attention due to their facinating thermal, electrical,
magnetic and transport properties,\cite{Nishino,kato,Nishino1,Nishino2} not only from the basic science prespectives
but also from the potential application as thermoelectric
materials.\cite{Nishino2,Nishino4,Mikami} Though Heusler-type
intermetallics commonly appear as metals,\cite{webster} semiconducting 
behaviour has been observed in Fe$_2$VAl and Fe$_2$VGa, as evidenced by their
negative temperature coefficient of resistivity\cite{Nishino}. This
 feature has been attributed to the appearance of a pseudo gap in
the density of states in the vicinity of the Fermi level, and these materials have been
characterized as semimetals. Nuclear magnetic resonance experiments  
and optical conductivity measurements further confirmed the existence of
deep pseudo gaps near the Fermi energy (E$_F$) in both 
compounds.\cite{Lue,Okamura} Recently, it has been reported by Lue et al.\cite{Lue1} that the
partial substitution of Ga by Ge in Fe$_2$VGa effectively dopes electrons to
the system thereby leading to a dramatic decrease in the electrical resistivity
A similar study for Fe$_2$VAl was carried out by Nishino et
al.\cite{Nishino3a} with the same conclusion. The variation of the Seebeck-coefficient with a sign
change from possitive to negative can be understood by means of a rigid band
like shifting of the Fermi level across the pseudogap. 

Several band structure
calculations have been reported providing various insights into Fe$_2$VAl and Fe$_2$VGa, all
confirming that Fe$_2$VAl is a non-magnetic semimetal with a pronounced pseudo gap
at the Fermi level\cite{Guo,Singh,Weht,Weinert,Bansil,Botton,Joo}. The calculated pseudo gap seems however too
wide to explain the experimental data. 
Guo et al.\cite{Guo} suggested the negative temperature coefficient of resistance to
arise from carrier localisation, while Singh and Mazin\cite{Singh} pointed towards the interaction of
carriers with fluctuating magnetic moments, and Weht and Pickett\cite{Weht} proposed dynamic
correlations between holes and electrons to be responsible for the
unusual resistivity behavior. The importance of magnetic moment formation in off-stoichiometric
compounds have been confirmed experimentally,\cite{Sumi} and recent theoretical studies of the
Fe$_{2+x}$V$_{1-x}$Z (Z=Al,Ga) compounds\cite{Denis,Fuji}
with a supercell approach predicted that antisite defects or excess atoms in the Heusler lattice may
induce a radical change in the electronic and magnetic properties. 
Similarly, Antonov et al.\cite{Antonov} have studied the electronic
structure and X-ray magnetic circular dichroism in Fe$_{2-x}$V$_{1+x}$Al for
various $x$ using first principles
calculations, which confirm the formation of magnetic moment on vanadium impurities. 

Alloys based on Fe$_2$VAl are good candidates for thermoelectric materials.\cite{Nishino2,Nishino4,Mikami} 
Although they have a high
thermoelectric power, because of their high electrical conductivities and high Seebeck coefficients, they
have poor figures of merit compared with other state-of-the-art thermoelectric
materials due to their high thermal conductivity, $\kappa$. Hence reduction of $\kappa$ is required for practical
applications,\cite{Mikami} which may be achieved by sustitution of Al by a heavier 
element.\cite{Nishino4} Large experimental efforts have been devoted to characterization of Fe$_2$VAl based
materials,\cite{Feng} including x-ray absorption\cite{Hsu} and photoemission\cite{Miyazaki,Soda} to reveal 
their surface and bulk electronic structure and electrical and thermal properties.\cite{Lue2,Nishino3,Kawaharada}
Yet, the lattice dynamics and
mechanical properties have not been explored, which is taken as the
objective of the present work. Here we present results of computations of the elastic constants and
phonon-dispersion relations for Fe$_2$VAl and
Fe$_2$VGa using density functional and density functional pertubation theory
within the generalised gradient approximation. From the modulus of elasticity,
we predict the mechanical behaviour of these compounds. 

\section {computational details}
%
The all-electron linear muffin-tin orbital method\cite{OKA} in a full-potential (FP-LMTO) 
implementation\cite{Savrasov} is used  to
calculate the total energies,  
and basic ground state and elastic properties of Fe$_2$VAl and Fe$_2$VGa.
In this method, the crystal volume is split   into two
regions: non-overlapping muffin-tin
spheres surrounding each atom and the
interstitial region between the spheres. We used a double $\kappa$
spdf LMTO basis to describe the valence bands, i.e. Hankel-functions characterized by decay constants $\kappa$ are smoothly augmented 
with numerical radial functions within the atomic spheres.
The calculations included the 4s, 4p, 3p, and 3d basis functions for iron and vanadium, 
the 3s, 3p and 3d basis for aluminum, and the 4s, 4p and 3d bases from gallium. 
The exchange correlation potential was 
calculated 
within 
the generalized gradient approximation (GGA) scheme\cite{Perdew}.
Inside the muffin-tin spheres, the charge density and potential  
were expanded in terms of spherical
harmonics up to $l_{max}$=6, while in the
interstitial region, they were expanded in plane
waves, with 12050 waves (energy up to 171 Ry) 
included in the calculation. Total energies
were calculated as a function of volume for a 
(28 28 28) k-mesh containing 624
k-points in the irreducible wedge of the Brillouin zone. The energy curves were fitted to a second order 
Birch equation of state\cite{Birch} to obtain the ground state properties.
The elastic constants were obtained from the variation of the total energy under 
volume-conserving strains, as outlined in Refs. \onlinecite{Oxides}.

In order to calculate the vibrational properties of the Heusler alloys  
we used the density functional perturbation theory as implemented in the plane-wave pseudopotential method, 
through the Quantum-ESPRESSO package\cite{pwscf}. We have used the Troullier-Martins 
non-conserving pseudopotentials\cite{TM}. Convergence tests lead to
 the choice of kinetic energy cutoffs of 80 Ry, and a (8 8 8) Monkhorst-Pack\cite{MP}
 grid of 
k-points for the Brillouin zone integration.


\begin{figure}
\label{figXRD}
\begin{center}
\includegraphics[width=100mm,angle=-90,clip]{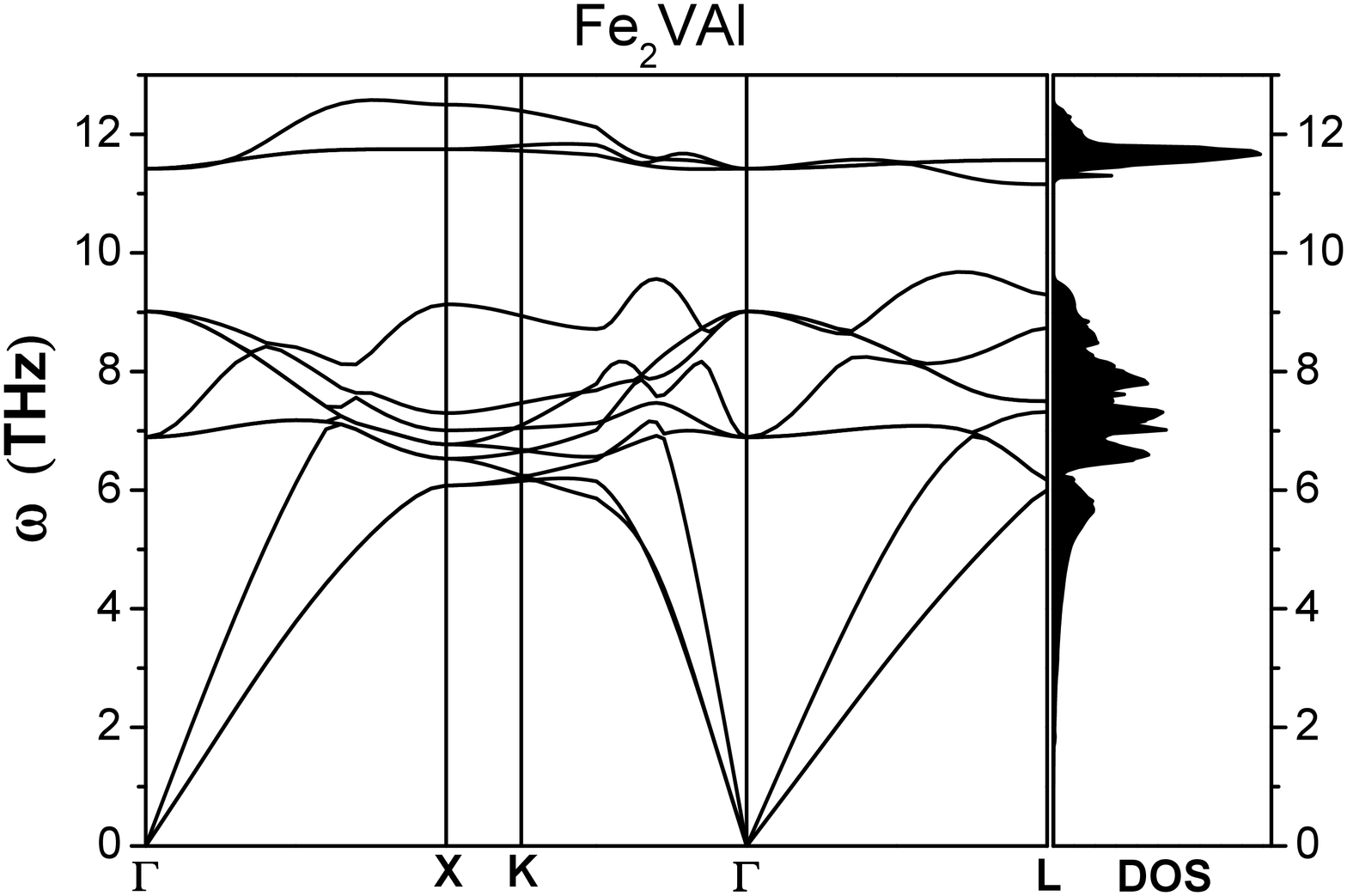}\\
\caption{Calculated phonon dispersion curves and phonon density of states of Fe$_2$VAl 
 }
\end{center}
\end{figure}


\section{phonons}
The calculated phonon dispersion curves of Fe$_2$VAl and Fe$_2$VGa are 
presented in figures 1 and 2, respectively. We do not find any imaginary phonon
frequency in the whole Brillouin zone for the two compounds. This supports the 
dynamical stability of both  compounds in the Heusler structure, which is not {\it a-priori}
evident in view of instabilities found in similar alloys, {\it e. g.}  Ni$_2$MnAl and  
Ni$_2$MnGa.\cite{Bungaro03,Zayak05} It is
interesting to note that the optical phonons for Fe$_2$VGa are coupled,
falling in a broad frequency range between  5.2 and 9.1 THz. This is in contrast to
Fe$_2$VAl, where three optical phonon branches are separated from the
lower frequency phonons. Inspection of the atomic masses of Fe, V, Ga and Al
allows us to understand this difference between the two
compounds. The similar atomic masses of Fe, V and Ga lead to the coupled phonon
dispersions, while the significantly lower atomic mass of Al results in the phonon
seperation. The three highest optical phonon modes of Fe$_2$VAl have their major amplitude 
contribution on Al atoms.


The lattice thermal conductivity is inversely proportional to the sound velocity,\cite{Holland} 
so assuming similar phonon 
scattering properties (relaxation times) for Fe$_2$VAl and  Fe$_2$VGa, 
the former compound thus should have the smaller thermal conductivity, which might be an observation of relevance
in efforts of minimization of thermal conductivity for good thermoelectric figures of merits. Note that in the
study of Ge substitution for Al in Fe$_2$VAl in Ref. \onlinecite{Nishino4} the opposite trend was found, i.e. 
$\kappa$ is reduced when the heavier Ge substitutes for Al, but this was attributed to the change in 
phonon relaxation time due to the incorporation of impurities with a large mass difference.

\section{Ground state and Elastic properties}
The calculated equilibrium lattice constant and the bulk modulus along with the experimental and 
other theory work is given Table. 1. The 
lattice constant obtained from the present calculation is underestimated by 0.8\% 
for Fe$_2$VAl and 0.7\% for Fe$_2$VGa when compared with 
the experimental value. The present results on lattice constant and bulk modulus 
agree quite well with the recent FLAPW calculations.\cite{Hsu}
The calculated elastic constants of Fe$_2$VAl and Fe$_2$VGa are presented in Table II. 
This is the first theoretical prediction of the elastic constants, 
which still awaits experimental confirmation. It is noticeable that the Cauchy pressure, $C_P\equiv C_{12}-C_{44}$ is negative
for Fe$_2$VAl ($C_P=-45$ GPa) but positive for Fe$_2$VGa ($C_P=+9$ GPa).
According to Pettifor\cite{Pettifor} this reflects a more covalent character of bonding in Fe$_2$VAl.
A simple
relationship, empirically linking the plastic properties of materials with their elastic 
moduli has been proposed by Pugh.\cite{Pugh} 
The shear modulus $G$ 
represents the resistance to plastic deformation, while the bulk 
modulus $B$ represents the resistance to fracture. A high $B/G$ ratio may then
be associated with ductility, whereas a low value would correspond to a more brittle nature. 
The critical value separating ductile and 
brittle materials is around $1.75$ according to Ref. \onlinecite{Pugh}. 
In the case of 
Fe$_2$VAl
the value of $B/G$ is $1.37$ and for Fe$_2$VGa the value is $1.73$ from our
calculated values in Table II,
and 
therefore Fe$_2$VAl can be classified as a brittle material, while Fe$_2$VGa 
resides on the borderline between brittle and ductile.  
Frantsevich \cite{fran} has a similar criterion based on the Poisson ratio, namely
$\nu < 1/3$ ($\nu > 1/3$) for brittle (ductile) character, hence
categorizing both compounds as brittle (see Table II). The difference 
merely reflects the imprecise nature of the concept.\cite{MgCNi3}

The sound velocities of Fe$_2$VAl and Fe$_2$VGa
may be derived from the calculated elastic constants,\cite{Oxides} see Table III. 
The properties of the two compounds are quite similar
apart from the heavier mass of Ga compared to Al, which is the dominating factor 
leading to smaller sound velocities in Fe$_2$VGa
compared to Fe$_2$VAl. The Debye temperature may be estimated from the simple isotropic 
approximations, $k_B \Theta_D=\hbar v_m k_D$, where
$k_B$ and $k_D$ are the Boltzmann constant and the Debye vector, respectively, and $v_m$ the average sound velocity. 
Hence, the smaller sound velocity
in Fe$_2$VGa directly leads to a smaller Debye temperature. 
The sound velocities may also be read from the acoustic phonon branches in figures 1 and 2. 
They are listed in the Table as well. One notices quite some anisotropy, the sound velocities along [110] 
being $\sim 30$ \% larger than
along the [100] and [111] directions.

The isomer shift of Fe in Fe$_2$VAl and Fe$_2$VGa have been measured by Ref. \onlinecite{Irwin}. The isomer shift
is proportional to the difference in electron contact density, $\rho(0)$, between the sample and reference
materials,
\[
I.S.=\alpha (\rho(0)-\rho_{ref}(0) ).
\]
We calculated
the electron contact density of Fe in the Fe$_2$VAl and Fe$_2$VGa 
compounds and compared to that of $\alpha$-Fe. 
Using the calibration
constant $\alpha=-0.22 a_0^3$ mm/s (Ref. \onlinecite{Eriksson}), this leads to the isomer shifts listed in Table IV. 
The difference in contact density is quite small, being larger in Fe$_2$VAl than in Fe$_2$VGa by 0.45 $a_0^{-3}.$ Thus
the isomer shift of the latter compound is larger by 0.10 mm/s, in excellent agreement with the experimental finding. 
However, the calculated isomer shifts relative to $\alpha$-Fe are about 0.04 mm/s smaller than measured, which could be
a temperature related effect, since the experiments are conducted at room temperature, while our theory pertains to 0 K.
The origin of the different value for the isomer shift in Fe$_2$VAl and Fe$_2$VGa 
is the difference in electronegativity of Al and Ga.
Although the electronic structures of these compounds are very similar,\cite{Bansil} 
the greater electronegativity of
Ga compared to Al transcripts into a deeper position of the Ga-dominated deepest $s$-band in Fe$_2$VGa 
compared to the position of the equivalent Al $s$-band in Fe$_2$VAl. This leads to reduced hybridization 
of Fe $s$ into this band in
Fe$_2$VGa compared to Fe$_2$VAl.  
To illustrate this, we plot in figure \ref{dos} the partial
density of states of Fe $s$-character for the two compounds. 
This is the relevant quantity to consider, since the dominant contribution to the
electron contact density {\it variations} in solids stems from the valence $s$-electrons (only $s$-states - and 
relativistic
$p_{1/2}$-states extend their amplitude into the nuclear region). Thus, in figure \ref{dos} the relevant
 $s$-band is the the lower part
of the density of states (between -10.3 eV and -7.5 eV in Fe$_2$VGa, between -9.3 eV and -6 eV in Fe$_2$VAl, 
relative to the Fermi level).
The reduced hybridization of Fe $s$ into this band leads to the smaller band width in Fe$_2$VGa compared to 
Fe$_2$VAl, as well as the
fewer integrated number of Fe $s$ electrons for this band (Table IV), which almost exactly reflects the 
difference in isomer shifts. 
In contrast, the upper
part of the Fe $s$ partial density of states in figure \ref{dos}, which describes the hybridization 
of Fe $s$ into the Fe and V $d$-bands,
 is much more similar in the two cases, and also integrates to almost the same number
of Fe $s$ charge. 
The larger hybridization in  Fe$_2$VAl compared to Fe$_2$VGa also explains the difference in sign of the Poisson ratio,
discussed above.


\begin{figure}
\begin{center}
\includegraphics[width=100mm,angle=-90,clip]{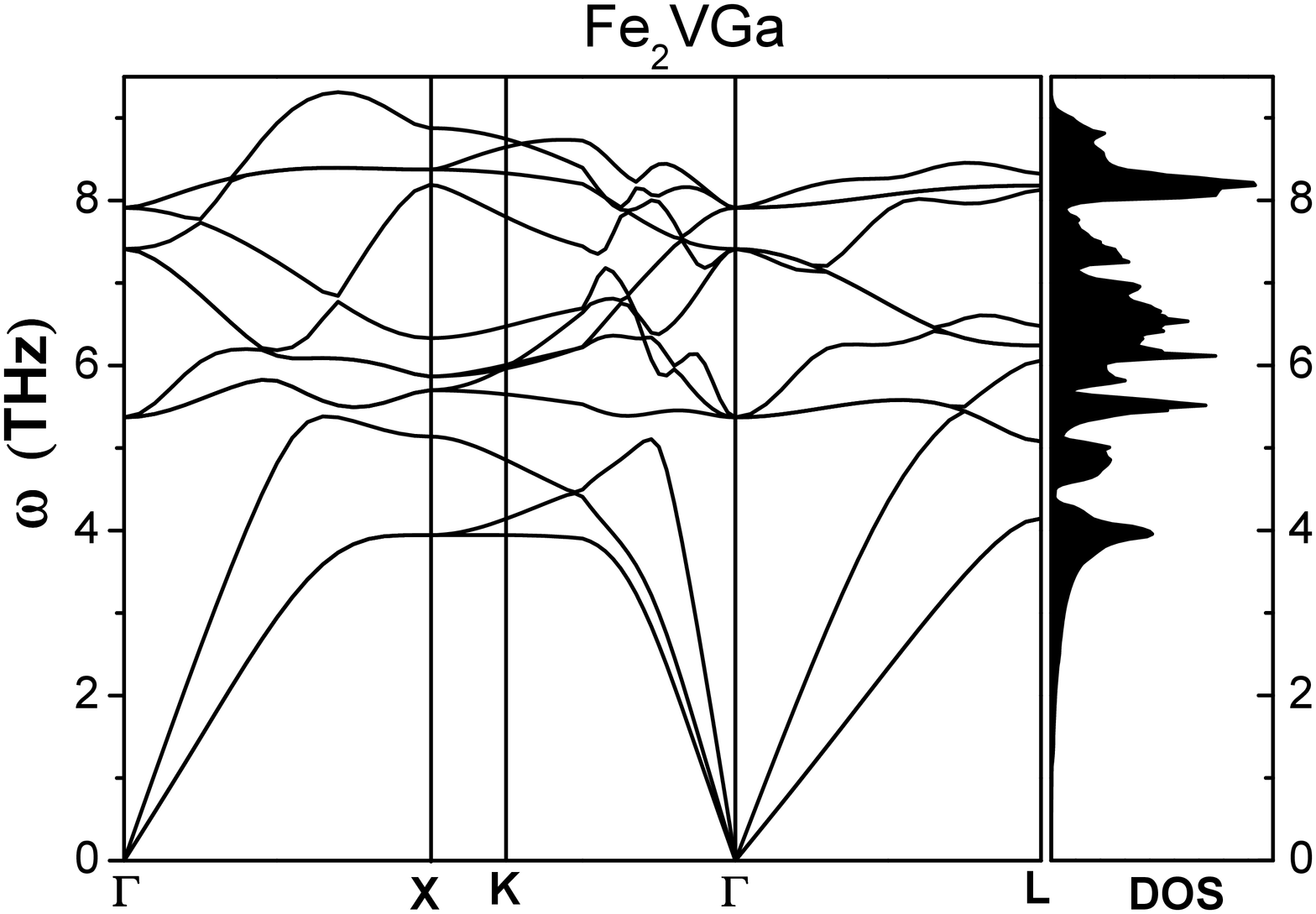}\\
\caption{Calculated phonon dispersion curves and phonon density of states of Fe$_2$VGa}
\label{EOS}
\end{center}
\end{figure}


\begin{figure}
\begin{center}
\includegraphics[width=100mm,angle=0,clip]{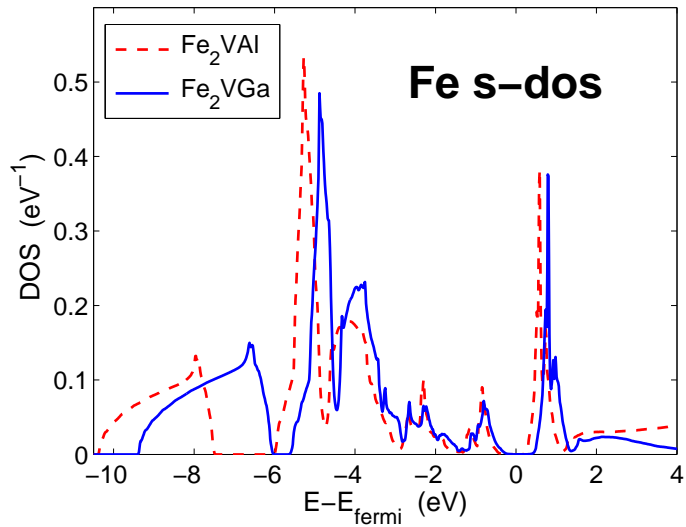}\\
\caption{Calculated Fe $s$ partial density of states (in units of electrons per eV and per Fe atom) 
in Fe$_2$VAl (red) and Fe$_2$VGa (blue). The zero of energy is placed at the 
Fermi level, which falls in the pseudo gap within the Fe and V $d$-bands.}
\label{dos}
\end{center}
\end{figure}

\section{Conclusions}

Ab initio electronic structure and phonon frequency 
calculations based on the density fucntional theory have been presented
for the Heusler alloys Fe$_2$VAl and Fe$_2$VGa. 
The elastic constants have been obtained, while no experiments exist for comparison. The differences
in phonon dispersions for the two compounds can be explained by the heavier mass of Ga compared to
Al, while the difference in Fe isomer shift is attributed to the higher electronegativity of Ga compared to Al, the
ensuing reduced covalency in Fe$_2$VGa compared to Fe$_2$VAl also explaining the difference in Cauchy pressure. 

\acknowledgments
G. V,  V. K acknowledge 
VR and SSF for financial support and SNIC for providing computer time. O.E 
acknowledges V.R for the support.
\clearpage

\begin{table}[tb]
\caption{
Calculated lattice constants in \AA, bulk modulus in GPa,
and its pressure derivative $B' $
of Fe$_2$VAl and Fe$_2$VGa at the theoretical equilibrium volume.}
\begin{ruledtabular}
\begin{tabular}{cccccc}
Compound   &      &Lattice Constant  & Bulk Modulus     & $B'$   \\
&&&&&\\
\hline
	
Fe$_2$VAl & Expt.        & 5.76                                     & -             & - \\
     & Theory, this work & 5.712                                    & 220.8                       &  5.3 \\
     & Other theory      & 5.712$^a$  & 212$^a$ &  -   \\

Fe$_2$VGa & Expt.        & 5.77                         & -              & - \\
     & Theory, this work & 5.727                         &228.5              & 4.9  \\
     & Other theory      & 5.726$^a$          &214$^a$ & -\\ 
     
\end{tabular}
\end{ruledtabular}
$^a$: Ref.\onlinecite{Hsu}; 
\end{table}

\begin{table}[tb]
\caption{
Calculated elastic constants, shear modulus (G), and Young's modulus (E) all expressed in GPa, 
and Poisson's ratio $\nu$
for Fe$_2$VAl and Fe$_2$VGa
at the theoretical equilibrium volume}.

\begin{ruledtabular}
\begin{tabular}{ccccccc}
              &$C_{11}$  &$C_{12}$  &$C_{44}$ &  G    &E      &$\nu$    \\ \hline

Fe$_2$VAl     & 415.7    & 125.3   &  170.7   & 160.5  & 387.6   & 0.208  \\ 
Fe$_2$VGa     & 363.1    & 161.0   &  152.3   & 131.8  & 331.7   & 0.258 \\ 
\end{tabular}
\end{ruledtabular}
\end{table}

\begin{table}[tb]
\caption{
Calculated longitudinal, shear, and average wave velocity ($v_l$, $v_s$, and $v_m$, respectively) 
in km/s, and the Debye temperature, $\theta_D$, in Kelvin from the average elastic wave velocity 
for Fe$_2$VAl and Fe$_2$VGa at the theoretical equilibrium volume. Also listed are the directional sound velocities
read from the acoustic phonon branches along directions [100], [111] and [110].}
\begin{ruledtabular}
\begin{tabular}{ccccccc}
 Compound    &     &\it{v$_l$}  &\it{v$_s$}  &\it{v$_m$} & $\theta_D$   \\ \hline

Fe$_2$VAl          & isotropic &  8.13    & 4.93 &  5.45 &  447    \\
                   & [100]   &  7.49    & 4.44 &       &         \\
                   & [111]   &  7.72    & 4.46 &       &         \\
                   & [110]   & 10.23    & 6.41 &       &         \\ 
Fe$_2$VGa          & isotropic &  7.09    & 4.05 &  4.50 &  369    \\
                   & [100]   &  6.29    & 3.52 &       &         \\
                   & [111]   &  6.26    & 3.13 &       &         \\
                   & [110]   &  8.15    & 4.52,4.19 &       &         \\ 
\end{tabular}
\end{ruledtabular}
\end{table}

\begin{table}[tb]
\caption{
Calculated and measured isomer shifts of Fe$_2$VAl and Fe$_2$VGa, 
in mm/s relative to $\alpha$-Fe. Also listed are the partial charges inside the Fe muffin-tin sphere
(radius $R=2.305$ a.u.). The Fe s-charge within the lowest $s$-like band is listed quoted in paranthesis (see text).}
\begin{ruledtabular}
\begin{tabular}{r|ll}
              &     Fe$_2$VAl  &  Fe$_2$VGa    \\ \hline
 I.S. theo    &       +0.021   &    +0.121     \\
 I.S. expt$^a$&       +0.058(5)&    +0.161(5)  \\ \hline
  Fe $n_s$    &        0.394(0.140)   &    0.370(0.111)      \\
     $n_p$    &        0.437   &    0.419      \\
     $n_d$    &        6.112   &    6.121      \\
     $n_f$    &        0.015   &    0.023      
\end{tabular}
\end{ruledtabular}
$^a$: Ref. \onlinecite{Irwin}
\end{table}

\end{document}